# Ultra-Thin Double-Walled Carbon Nanotubes: A Novel Nanocontainer for Preparing Atomic Wires

Lei Shi[1], Leimei Sheng[1] (✉), Liming Yu[1], Kang An[1], Yoshinori Ando[2], and Xinluo Zhao[1] (✉)

[1] Department of Physics, and Institute of Low-Dimensional Carbons and Device Physics, Shanghai University, Shanghai 200444, China
[2] Department of Materials Science and Engineering, Meijo University, Nagoya 468-8502, Japan

## ABSTRACT

Double-walled carbon nanotubes (DWCNTs) with high graphitization have been synthesized by hydrogen arc discharge. The obtained DWCNTs have a narrow distribution of diameters of both the inner and outer tubes, and more than half of the DWCNTs have inner diameters in the range 0.6–1.0 nm. Field electron emission from a DWCNT cathode to an anode has been measured, and the emission current density of DWCNTs reached 1 A/cm² at an applied field of about 4.3 V/μm. After high-temperature treatment of DWCNTs, long linear carbon chains (C-chains) can be grown inside the ultra-thin DWCNTs to form a novel C-chain@DWCNT nanostructure, showing that these ultra-thin DWCNTs are an appropriate nanocontainer for preparing truly one-dimensional nanostructures with one-atom-diameter.

## 1. Introduction

The discovery of carbon nanotubes (CNTs) heralded the beginning of the boom of nanoscience and nanotechnology [1], and the successful synthesis of graphene, a superlative two-dimensional carbon nanomaterial, has promoted further extensive exploration in the nanoworld [2]. The question is: what is next? Recently, atomic-scale nanowires have attracted more and more interest due to their truly one-dimensional (1D) nanostructures with one-atom-diameter [3–5]. Generally, atomic-scale nanowires are not stable in air at room temperature. One of the methods to preserve them is to encapsulate them into CNTs. The innermost diameters of these CNTs are required to be in the range 0.6–1.0 nm in order to ensure that linear atomic wires can be formed [3–6]. Double-walled carbon nanotubes (DWCNTs), consisting of two concentric graphene cylinders, are the simplest member of the family of multi-walled carbon nanotubes (MWCNTs), and have even better electronic, thermal and mechanical properties than single-walled carbon nanotubes (SWCNTs) [7–10]. Therefore, DWCNTs have been used as a novel nanocontainer for preparation of 1D materials [4–6].

Many attempts have been made to synthesize high purity DWCNTs, such as conventional arc discharge [11–13], high-temperature pulsed arc discharge [14], high-temperature annealing of peapods [15], and chemical vapor deposition (CVD) [16–19]. Furthermore,

Address correspondence to Leimei Sheng, shenglm@staff.shu.edu.cn; Xinluo Zhao, xlzhao@shu.edu.cn

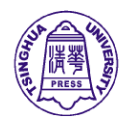

density gradient ultracentrifugation (DGU) has been also used to separate DWCNTs from SWCNTs [20]. Arc discharge is a very important technique to produce fullerenes, MWCNTs, SWCNTs, and graphene [1, 21–23]. However, the DWCNTs prepared by arc discharge possess large average inner diameters (> 2 nm), even though they have the highest crystallinity due to having the highest growth temperature among the methods mentioned above [11–13].

Herein, we report a simple method for mass-production of high purity, uniform, and ultra-thin DWCNTs with high graphitization by hydrogen arc discharge. More than half of the DWCNTs have inner diameters in the range 0.6–1.0 nm, and long linear carbon chains (C-chains) can be grown inside the ultra-thin DWCNTs to form a new 1D carbon nanostructure, C-chain@DWCNT, by high-temperature treatment. Our results will not only enable investigation of the electronic, thermal, and mechanical properties of ultra-thin DWCNTs with high crystallinity, but also provide a new nanocontainer for synthesizing linear atomic wires, such as C-chain, a novel 1D allotrope of carbon composed of sp-hybridized carbon atoms.

## 2. Experimental

A home-made apparatus featuring dc arc discharge evaporation was used to synthesize DWCNTs, where two electrodes were installed horizontally. The anode (6 mm diameter) was prepared from a mixture of 96 wt.% graphite powder and 4 wt.% Fe catalyst. The cathode (10 mm diameter) was a pure carbon electrode. A dc arc discharge was generated between these two carbon electrodes by applying 75 A in an atmosphere of Ar and $H_2$ mixture (3:2/*v:v*) at 300 torr. After 30 min of arc evaporation, over 300 mg of as-grown CNTs (SWCNTs, DWCNTs, and three-walled CNTs) were obtained. The as-grown CNT sample was purified by a three-step process: Oxidation in air at 400 °C for 30 min, and then at 500 °C for 2 h, and finally immersion into concentrated hydrochloric acid for 12 h at room temperature. In the end, a DWCNT buckypaper could be obtained after washing with distilled water and filtration using a polytetrafluoroethylene filter.

The samples were characterized by scanning electron microscopy (SEM) (JEOL JSM-6700F) equipped with an energy dispersive spectroscopy analyzer (EDS, Oxford Inca), transmission electron microscopy (TEM) (JEOL JEM-200CX), high-resolution TEM (HRTEM) (Tecnai G2 F20 S-Twin), Raman spectrometry (Renishaw, inVia Plus and Horiba Jobin Yvon, T64000), and thermogravimetric analysis (TGA) (Netzsch, TG209F1). Field emission measurements were carried out in a vacuum chamber with a base pressure of $3 \times 10^{-5}$ Pa (Keithley 2410 SourceMeter).

## 3. Results and discussion

Figures 1(a) and 1(b) show typical SEM and TEM images of DWCNTs. A compact structure is formed due to the increase in the attractive van der Waals force between the DWCNTs after the removal of the Fe catalyst and other carbon impurities in the three-step purification process. A representative HRTEM image is shown in Fig. 1(c). All the CNTs that can be distinguished are DWCNTs, and about ten isolated DWCNTs with similar inner diameters of around 0.9 nm can be seen in this image. Examination of about one hundred such HRTEM images reveals that our sample consisted mainly of DWCNTs (over 95%, Fig. 1(d)). The diameter distribution of the DWCNTs showed inner and outer diameters in the range 0.4–1.8 nm and 1.2–2.6 nm, respectively (Fig. 1(e)). In particular, more than half of the DWCNTs had inner diameters in the range 0.6–1.0 nm, and should be appropriate nanocontainers for linear atomic wires. The average values of the inner and outer diameters of the DWCNTs were 0.9 nm and 1.6 nm, respectively.

Generally, there are two reasons which can explain why DWCNTs, but not other CNTs, were synthesized using our method: the size of catalyst nanoparticles and the relative concentration of active carbon species [11–13, 17]. The catalysts used in the synthesis of DWCNTs by arc discharge usually include sulfur and elements in the iron group in the anode [11–13]. Sulfur has long been used to synthesize carbon fibers, and its effects on the amount and quality of carbon products have been discussed in detail [24, 25]. The optimum sulfur content can improve the carbon deposition on metal catalyst nanoparticles and promote the growth of CNTs; high contents of S in the anode will make the catalyst nanoparticles aggregate into somewhat larger nanoparticles, resulting in the formation of thick CNTs.

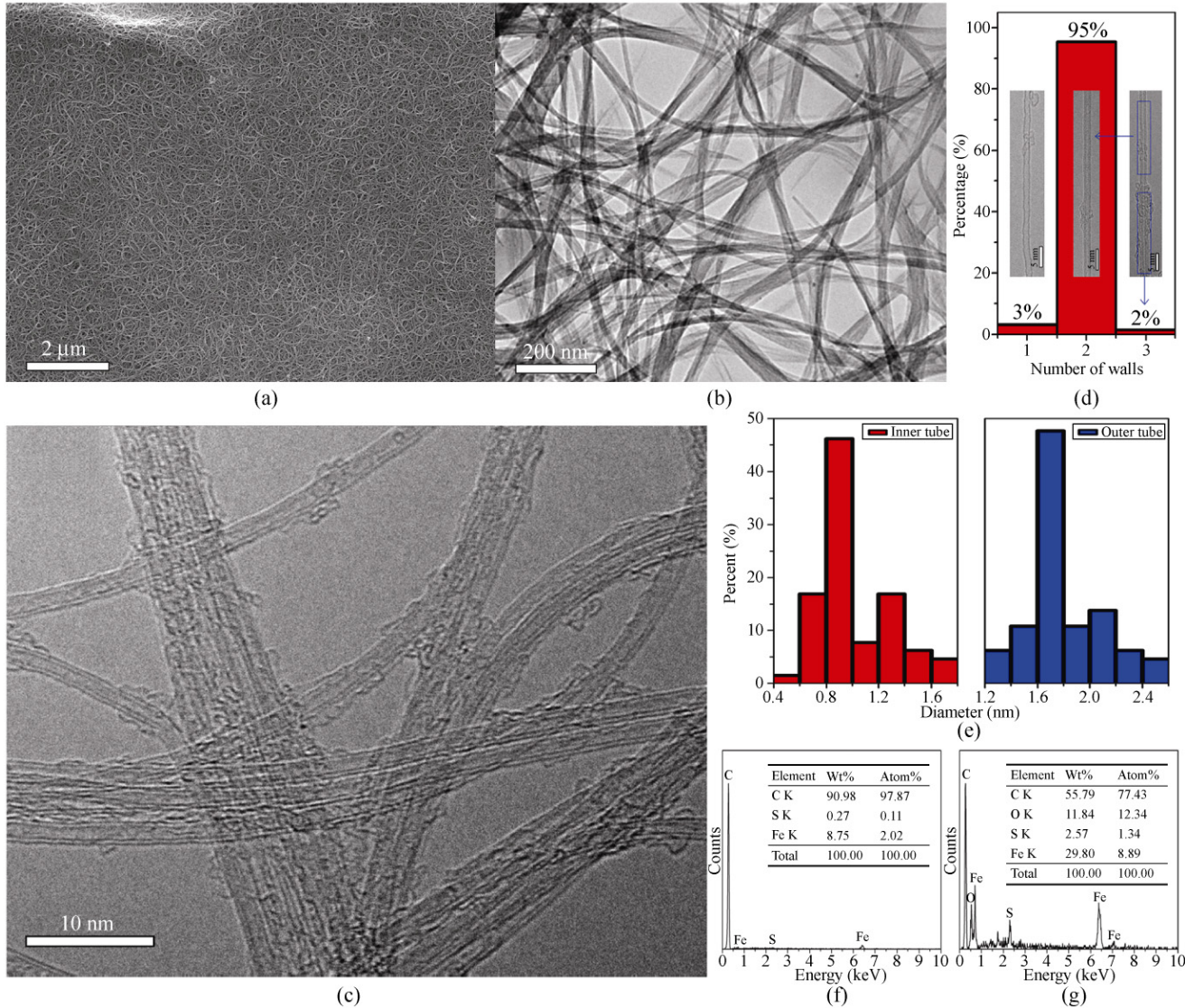

**Figure 1** (a) SEM, (b) TEM, and (c) HRTEM images of DWCNTs; (d) ratio of single-, double-, and multi-walled CNTs, with the insets showing HRTEM images of the different CNTs; (e) diameter distribution of DWCNTs based on detailed HRTEM observations; EDS of the anode (f), and as-grown DWCNTs (g)

In our experiments, the DWCNT synthesis does not require additional S catalyst, although in actual fact, about 0.3 wt.% S was present in the anode because the raw carbon materials contained a little sulfur, and the high-temperature treatment for preparing the anode could not remove all of this element (Figs. 1(f) and 1(g)). The high yield of uniform and ultra-thin DWCNTs can be attributed to the low content of S in the anode. In addition, the appropriate current, atmosphere and pressure are also the reasons for being able to synthesize ultra-thin DWCNTs with high yield, because these parameters determine the amount of active carbon species used to form CNTs as well as the reaction time that the CNTs stay in the reaction zone. We believe that enhancing the feed rate of active carbon species and prolonging the reaction time both benefit the formation of DWCNTs rather than SWCNTs.

The detailed features of the DWCNTs, such as purity, structural integrity, diameter distribution, and nanotube type, were examined using Raman scattering. Five laser lines at 488, 514, 532, 633, and 785 nm were used for excitation (Fig. 2) in order to give a general diameter distribution, since different diameter tubes are in resonance with different laser energies. A well-known fact is that the outer and the inner tubes of a DWCNT can be either metallic (M) or semiconducting (S). The use of different laser lines allowed us to study different configurations of outer/inner tubes separately, such as M/S (outer tube type/inner tube type) and S/M DWCNTs. For ease of comparison, Raman spectra have been normalized to the intensity of the G-band, and the radial breathing modes (RBMs) were expanded by a factor of 10 except for 785 nm excitation. In the high frequency region of the Raman spectra, the position of the D-band showed a blue shift (from 1313 to 1350 cm$^{-1}$) with increasing laser energy. The intensity of the D-band was very low, with the intensity ratio of the D-band to the G-band ($I_D/I_G$) in the region of 0.02–0.04, meaning that the DWCNT samples have high crystallinity and relatively low amounts of amorphous carbon (Table 1).

**Table 1** Diameter analysis of DWCNTs using Raman spectra recorded at different excitation wavelengths

| Wavelength (nm) | RBM peak (cm$^{-1}$)/ Diameter (nm) Outer tube | RBM peak (cm$^{-1}$)/ Diameter (nm) Inner tube | Interlayer spacing (nm) | $I_D/I_G$ |
|---|---|---|---|---|
| 785 | 158/1.58 | 264/0.92 | 0.66 | 0.038 |
| 633 | 126/2.02 | 200/1.23 | 0.79 | 0.021 |
|  | 151/1.66 | 253/0.96 | 0.70 | — |
| 532 | 158/1.58 | 268/0.91 | 0.67 | 0.026 |
| 514 | 153/1.64 | 262/0.93 | 0.71 | 0.023 |
| 488 | 126/2.02 | 200/1.23 | 0.79 | 0.026 |
|  | 135/1.87 | 231/1.06 | 0.81 | — |
|  | 155/1.61 | 265/0.92 | 0.69 | — |
|  | 166/1.50 | 302/0.80 | 0.70 | — |

We estimated the diameters of DWCNTs using the relation

$$\omega_{RBM} = 234/d_t + 10 \text{ cm}^{-1}$$

from the RBM peaks in the low-frequency range from 100 to 350 cm$^{-1}$, where $\omega_{RBM}$ (cm$^{-1}$) is the RBM frequency and $d_t$ (nm) is the nanotube diameter (Table 1) [26]. The RBM bands of DWCNTs clearly indicate the coexistence of inner and outer tubes. According to the relation, the strongest RBMs of the outer tubes (153–158 cm$^{-1}$) correspond to diameters of 1.58–1.64 nm, and the diameters of the inner tubes with the strongest RBMs at 262–268 cm$^{-1}$ are 0.91–0.93 nm (Fig. 2). The difference between the outer and inner diameters is 0.65–0.73 nm, which is twice the lattice spacing of graphite (002), as expected. The Raman data are consistent with the HRTEM observations. The Raman and HRTEM results suggest that the DWCNTs have a narrow diameter distribution. It is very interesting that the RBM peaks of the DWCNTs prepared by our method are almost the same as those of the DWCNTs synthesized by the CVD method of Endo et al. [16], and by the DGU technique [20]. Furthermore, all of these samples have another common characteristic, that is, higher purity of DWCNTs than other samples previously reported in the Ref. [11–14, 17, 18].

According to the "Kataura" plot, $E_{laser}$ = 1.58 eV (785 nm) is in resonance with $E_{11}^M$ and $E_{22}^S$ for the outer and inner tube, respectively [27]. The strongest peak located ~264 cm$^{-1}$ is associated with semiconducting tubes belonging to the $2n + m = 22$ family and resonant with $E_{22}^S$ transitions. This laser energy is useful to detect the M/S (outer/inner) DWCNTs, while $E_{laser}$ = 2.41 and 2.33 eV (514 and 532 nm) are selective for the S/M DWCNTs, which are in resonance with $E_{33}^S$ (or $E_{44}^S$) and $E_{11}^M$ for the outer and inner tubes, respectively. Similarly, for $E_{laser}$ = 1.96 and 2.54 eV (633 and 488 nm), are resonant with S/S and M/M DWCNTs, respectively. Obviously, our DWCNT sample provides an appropriate candidate to clearly study the different types (S or M) of outer and inner tubes by using different Raman laser energies.

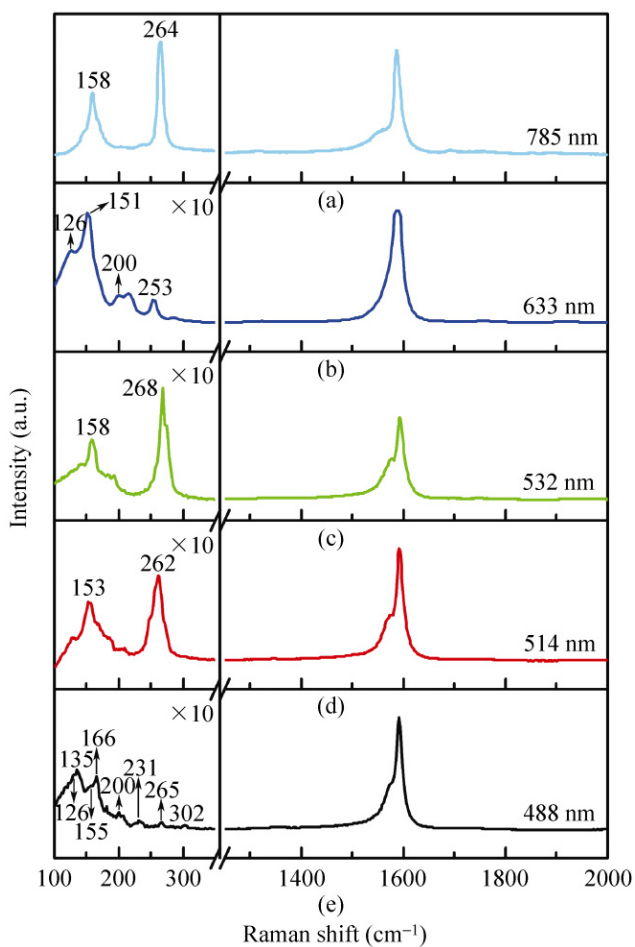

**Figure 2** Raman spectra of DWCNTs with excitation wavelengths of 488 nm, 514 nm, 532 nm, 633 nm, and 785 nm

To evaluate the purity and the quality of DWCNTs, TGA was conducted in air with a heating rate of 10 °C min$^{-1}$ using a thermogravimetric analyzer. The residue was less than 3 wt.% after heating to 950 °C, confirming that the DWCNT sample has a high purity (Fig. 3(a)). The two peaks of the differential thermogravimetric (DTG) curve correspond to the burning of SWCNTs and DWCNTs, in ascending order of temperature. The DTG peak corresponding to DWCNTs indicates that the DWCNTs have an excellent high-temperature oxidation resistance up to 786 °C, which is the highest burning temperature of DWCNTs to our knowledge.

After a three-step purification process, a DWCNT "buckypaper" could be obtained (Fig. 3(b)). This "buckypaper" has a diameter of ~20 mm, and is tough and flexible enough to fold. Therefore, our method provides a simple and effective way to mass-produce high purity, uniform and ultra-thin DWCNTs with high crystallinity. The high quality of DWCNTs has relationship with the low S content, the high growth temperature, and the defect-healing effect by hydrogen arc discharge method.

To demonstrate the excellent electric and thermal characteristics of our uniform and thin DWCNTs, field electron emission from a DWCNT cathode to an anode was measured. A piece of DWCNT buckypaper was directly used as emitting cathode. As shown in Fig. 4, the turn-on and threshold electric fields of DWCNTs, corresponding to a current density of 1 μA/cm$^2$ and 1 mA/cm$^2$, are about 0.9 and 2.0 V/μm, respectively. The emission current density of DWCNTs reached 1 A/cm$^2$ at an applied field of about 4.3 V/μm. The inset (a) of Fig. 4 shows the Fowler–Nordheim

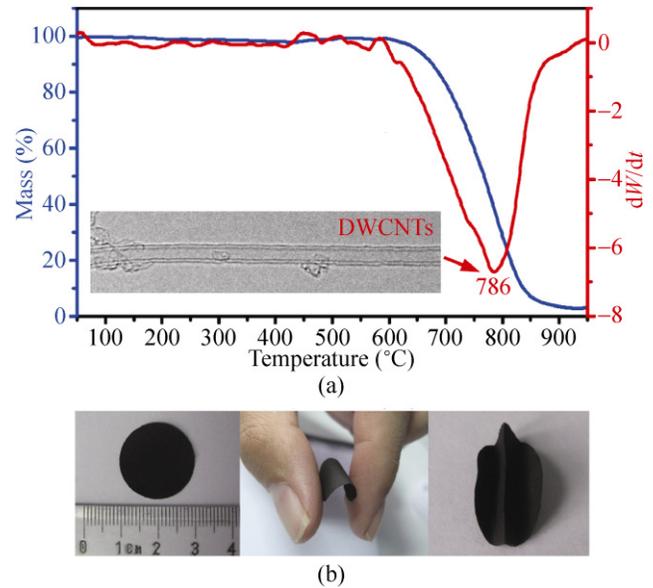

**Figure 3** (a) TGA and DTG curves of DWCNTs, with the insets showing HRTEM image of DWCNTs; (b) photographs of DWCNT buckypaper, which is tough and flexible enough to fold

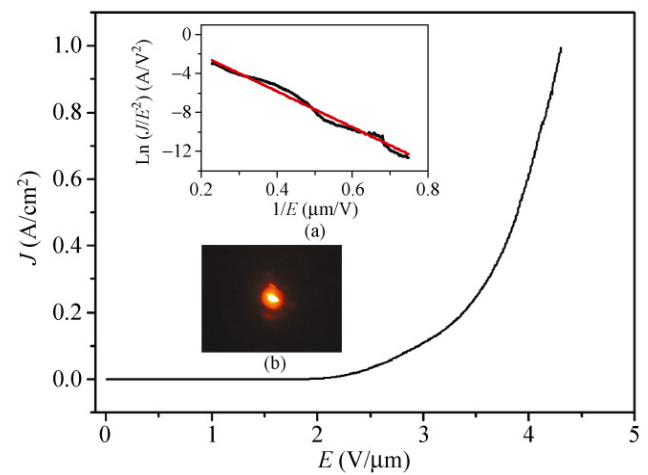

**Figure 4** Field emission properties of DWCNTs. Inset (a) shows an F–N plot with the slope indicated by a red straight line; inset (b) is an anode luminescence image with $E$ = 2.4 V/μm

(F–N) plot and its fitting line, and the inset (b) in Fig. 4 is the anode luminescence image at 2.4 V/μm. The field enhancement factor can be calculated using the relation obtained from the F–N equation [28]

$$\beta = -B\Phi^{3/2}/S_{FN},$$

where $\beta$ is the field enhancement factor, $B = 6.83 \times 10^9$ eV$^{-3/2}$ Vm$^{-1}$, $\Phi$ is the work function, and $S_{FN}$ is the slope of the fitting line. Assuming the work function of CNTs to be 5.0 eV, the field enhancement factor is estimated to be 3646. By comparison with that of DWCNTs having a wide diameter distribution [29], the uniform, ultra-thin, and high-purity DWCNTs with high crystallinity possess much lower turn-on field, higher emission current density, and higher field enhancement factor.

Linear C-chains are a truly 1D carbon nanomaterial, but they cannot survive in air. A carbon nanowire consisting of a MWCNT with a long 1D linear C-chain inserted into its innermost tube of 0.7 nm in diameter has been discovered in the cathode deposits prepared by hydrogen arc discharge [3]. It has been proposed that the CNTs used to confine linear C-chains must have inner diameters of around 0.7 nm. Since in our sample more than 15% of the DWCNTs have diameters of 0.6–0.8 nm (Fig. 1(e)), we carried out a high-temperature treatment on our DWCNTs under an Ar atmosphere at a temperature of 1500 °C for 30 min. Figure 5 shows the Raman spectra of DWCNTs and the heat treated DWCNTs (HT-DWCNTs). The Raman band at 1856 cm$^{-1}$ (the corresponding second order Raman peak is located at 3693 cm$^{-1}$) clearly seen in the inset of Fig. 5, is considered to be characteristic of long linear C-chains inside CNTs [3], and indicates the formation of a new 1D carbon nanostructure, C-chain@DWCNT. Endo et al. attributed a Raman peak at 1855 cm$^{-1}$ to the presence of short carbon chains (e.g., 3–7 atoms long), linked covalently to adjacent tubes after high-temperature treatment of highly pure DWCNTs [30]. We suggest that the band at 1856 cm$^{-1}$ in the spectrum of our heat treated material can be taken to indicate the formation of long linear C-chains from carbon atoms contained inside the inner tubes of the DWCNT sample after high-temperature treatment. It is known that there are some carbon atoms existing inside the inner tubes of DWCNTs. During the high-temperature treatment process, these carbon atoms move along the tube and gradually aggregate into clusters in the case of larger diameter DWCNTs, which cannot confine the carbon atoms into a line. If the DWCNTs have inner diameters of around 0.7 nm, however, a long linear C-chain will be formed inside. In fact, the Raman band at around 1855 cm$^{-1}$ is not observed after heat-treating the large diameter SWCNTs or DWCNTs, which provides indirect confirmation of the formation mechanism. Further detailed studies of the mechanism are in progress in our laboratory.

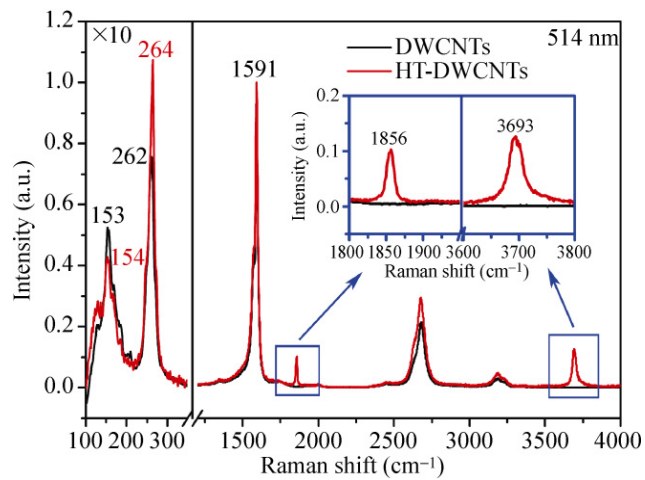

**Figure 5** Raman spectra of DWCNTs and HT-DWCNTs with an excitation wavelength of 514 nm

## 4. Conclusions

Uniform and ultra-thin DWCNTs with high graphitization and high purity have been mass-produced by a simple hydrogen arc discharge without the need for an additional sulfur catalyst. The DWCNT cathode has excellent field emission properties and the emission current density of the DWCNT buckypaper can reach 1 A/cm$^2$ at about 4.3 V/μm. It has been confirmed that long linear C-chains can be grown inside the ultra-thin DWCNTs to form a novel C-chain@DWCNT nanostructure by high-temperature treatment. Both the ultra-thin DWCNTs and C-chain@DWCNT nanostructure are 1D materials which are candidates for use in the fabrication of, for example, nanocomposites, field emission sources and nanodevices.

## Acknowledgements

This work was supported by the National Natural Science Foundation of China (Grant No. 10974131), the Nanotechnology Program of Shanghai Science and Technology Committee (No. 0952nm07100), the Science and Technology Innovation Fund of the Shanghai Education Committee (No. 09ZZ85) and Shanghai Pujiang Talent Plan (No. 08PJ1405100). We thank Professor Shoushan Fan of Tsinghua–Foxconn Nanotechnology Research Center for the HRTEM measurements, and Professor Pingheng Tan of the Institute of Semiconductors for some of the Raman measurements.

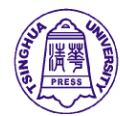